# Acoustic Sensing after 50 km of Transmission Fibre using Coherent Optical Subassembly

Florian Azendorf[(1)], André Sandmann[(1)], Wolfgang Reimer[(1)], Michael Eiselt[(1)]

[(1)] Adtran, Märzenquelle 1-3, 98617 Meiningen, Advanced Technology, fazendorf@adva.com

**Abstract**
A coherent optical subassembly (COSA) is evaluated for coherent-correlation optical time domain reflectometry (CC-OTDR) based fibre sensing. Even though the COSA was originally designed for digital communication applications, acoustic signals with frequencies up to 360 Hz can be detected after 50 km of transmission. ©2023 The Authors

**Introduction**

Optical time domain reflectometers (OTDRs) are proven devices to ensure quality of transmission in optical fibre networks. Thanks to their high sensitivity, spans of more than hundred kilometres can be monitored. In addition to measuring fibre attenuation, they can be used to detect splices and bends along the fibre link. However, their sensitivity in terms of power changes is not sufficient to detect weak fluctuations in the Rayleigh backscattering caused by environmental effects in the proximity of the fibre. In order to facilitate detection of these weak variations and dynamic events in the backscattered trace, substituting the direct detection scheme by coherent detection turned out to be a valid approach. Through coherent detection, it is possible to evaluate amplitude, phase, and polarization of the backscattered signal. The phase information is highly sensitive to dynamic fluctuations such as vibrations in the fibre environment. In addition, the signal-to-noise ratio (SNR) can be increased by sending a long pulse sequence instead of a single short pulse. In order to obtain narrow reflection peaks, the backscattered signal is cross-correlated with the transmitted sequence.

In the last years, several studies were published using a Coherent OTDR (C-OTDR) to monitor fibre network infrastructure in Texas [1], New Jersey [2], and France Nozay [3]. In all these references, the authors investigated the impacts of fibre sensing on data transmission in the conventional wavelength band (C-band). The results of these publications show that sensing the environment is compatible with simultaneous data transmission. Although linear cross-talk is not observed as the sensing channel has a small spectral width, high powers should be avoided in the sensing channel to prevent non-linear cross-talk on channels in dense wavelength division multiplexed networks. In these studies, commercially available sensing systems were used [1, 2]. Such systems use high-performance optical components that are optimized for sensing applications and come at high cost. Nowadays, integrated transceivers are available in the telecommunication industry with significantly lower costs due to the high production volumes of these components.

We evaluate the operation of a coherent correlation OTDR (CC-OTDR) using a coherent optical subassembly (COSA) with an external laser input optimized for telecommunication applications [4]. Acoustic signals are detected in the Rayleigh backscattering trace after 50 km of standard single-mode fibre (SSMF) over a frequency range from 196 Hz to 360 Hz. In a second experiment, the reflections from two physical contact connectors with 10 m separation at the end of a 70 km fibre are used for audio detection. In this experiment, audio frequencies in the range from 200 Hz to 625 Hz were successfully detected. We thereby show that the COSA is an appropriate device for environmental sensing.

**Experimental Setup**

The experimental CC-OTDR testbed with the COSA is schematically illustrated in Fig. 1. Here, the continuous wave signal of a highly coherent laser with a Lorentzian linewidth of less than 100 Hz is fed into the local oscillator (LO) input of the COSA. A coupler splits the optical power of the laser signal into two parts. Subsequently, one part of the light is sent to the coherent receiver and acts as the LO. The phase of the other part is modulated at bit rates of 400 Mbit/s and 250 Mbit/s using binary phase shift keying (BPSK). The transmitted code is a pseudo-random binary sequence (PRBS) with a length of 8191 plus a trailing '-1' symbol to obtain a balanced signal. Only a single channel in one of the orthogonal polarizations (x-polarization) is used. The other channels are not connected with the arbitrary waveform generator and their modulator biases are set to full destructive interference. The probing signal frame consists of the PRBS plus a fill pattern, which is needed to suppress the modulator output between the PRBS bursts. The repetition rate of the probing signal frame determines the maximum detectable frequency of a dynamic event. An erbium-doped fibre amplifier (EDFA) boosts the probing signal, and the amplified spontaneous emission is filtered by a subsequent optical bandpass filter (OBF). Afterwards, the probe signal is coupled into the fibre via an optical circulator. The signal backscattered or reflected from the fibre is received in the coherent receiver and recorded on four channels of a real-time oscilloscope. In order to mitigate aliasing, low-pass filters (LPF) are placed between the COSA and the real-time oscilloscope. Each received signal is cross-correlated with the transmitted PRBS signal, and the squared sum of the correlations, representing the reflected power, is displayed on the bottom right of Fig. 1. The three red dots visible in the trace mark the reflections at the sensor fibre and the patch cord connectors. Here, the reflections at the input and the output of the 10-meter long sensor fibre are caused by two LC/PC connectors. Finally, an angled connector with a dust cap causes the last reflection. The sensor fibre is wrapped up in a loose tube buffer and a speaker is used to induce dynamic events at its last two meters. Signal frames are recorded over a duration of 50 milliseconds and from each frame a sample of the propagation phase between neighbouring scattering or reflection points is calculated. The spectrum of the phase variations is obtained by the fast Fourier transform and compared to the frequency of the applied acoustic signal.

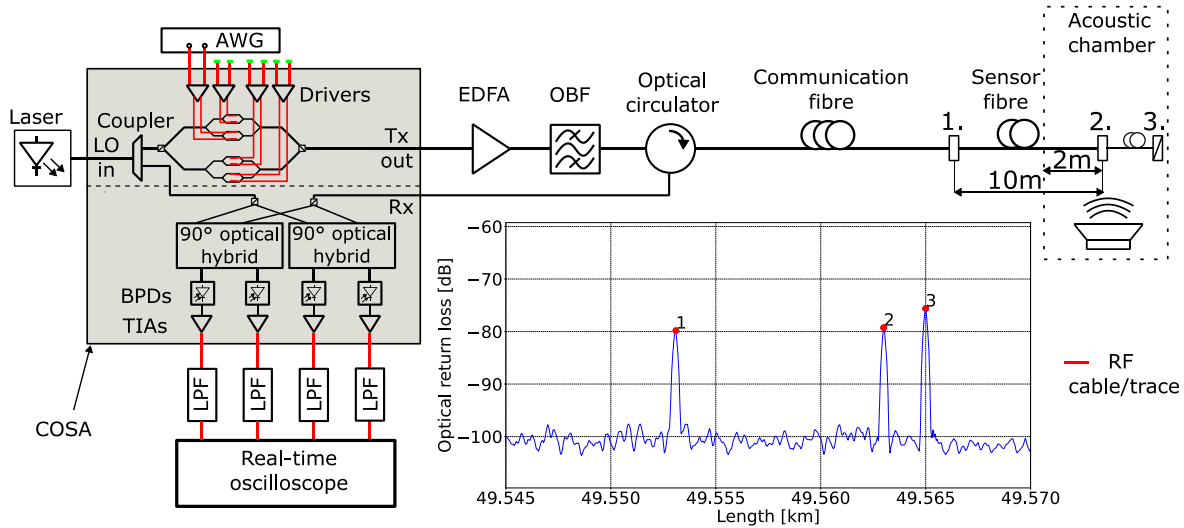

**Fig. 1:** Schematic of the experimental setup. BPD: balanced photo diode, TIA: transimpedance amplifier

**Results obtained with 50 km fibre**

In the first experiment, the SSMF has a length of 50 km. A patch cord with an open APC connector covered by a dust cap is connected to the output of the sensor end to avoid saturation effects at the receiver. In Fig. 2, the fingerprint (optical backscatter trace) of the concatenated fibres is shown with reflection peaks at the connector locations. The input reflection with the highest amplitude was referenced to -55 dB, which was obtained in a calibration measurement with an optical power meter. Additionally, the round-trip fibre attenuation is determined from the envelope slope in the figure to equal 0.38 dB/km, which corresponds to a single-pass attenuation of 0.19 dB/km. The thermal noise floor level in Fig. 2 has an average value of -103 dB, while the side lobes level, caused by the cross-correlation of the PRBS with the received signal, has a discrete distance to the main lobe of 38 dB in this measurement, the level is, on average, at -100 dB.

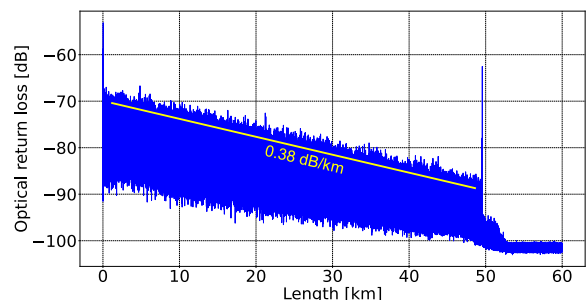

**Fig. 2:** Averaged fingerprint of the communication fibre

Fig. 3 illustrates the averaged fingerprint of a 25-meter fibre section after approximately 50 km of communication fibre with three reflection peaks from connectors, as shown in Fig. 1. Additionally, the highest eight Rayleigh backscatter peaks are marked with red dots and numbered accordingly. These peaks originate from the coherent superposition of the scattered electromagnetic fields from different locations along the fibre, leading to an interference pattern. The phase variations caused by the acoustic signal of the speaker are evaluated by calculating the phase difference between two backscatter peaks. Fig. 4 shows the normalized spectra exemplarily for the phase differences between peak 4 and peak 8. In a second experiment with the 50 km SSMF, the sampling rate is changed from 2 GS/s to 1 GS/s. Here, the decrease in sampling rate leads to an increase in the number of recorded frames and hence to an improved frequency resolution from 39 Hz to 20 Hz. In addition, the SNR is improved by reducing the baud rate from 400 MBaud to 250 MBaud. Correspondingly, the probe signal durations are 540,5 µs and 632,8 µs, leading to a maximum detectable frequency of 925 Hz and 790 Hz, respectively. Furthermore, the distance between peaks 4 and 8 is 5 m leading to a gauge length in the same order of magnitude. The frequencies of the applied sinusoidal signals are clearly visible in the spectra at 195 Hz, 280 Hz, and 360 Hz, respectively, as shown in Fig. 4.

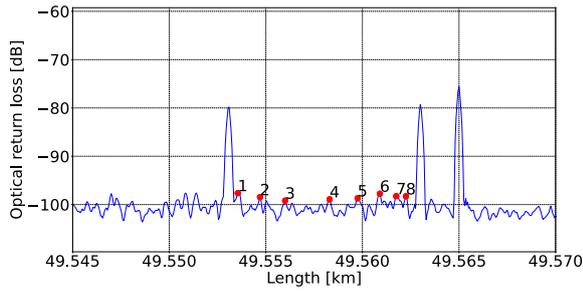

**Fig. 3** Correlation peak maxima in the backscattering trace when a 270 Hz acoustic signal was applied

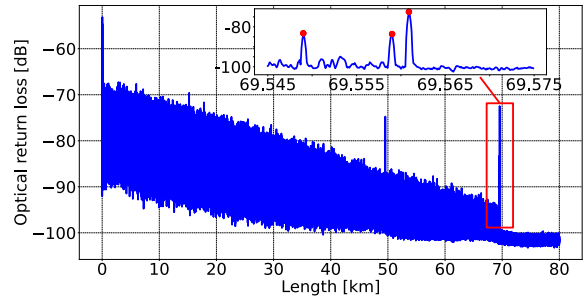

**Fig. 5** Averaged fingerprint over 70 km of communication fibre with inset showing the last three reflections of the fibre connectors and adapters

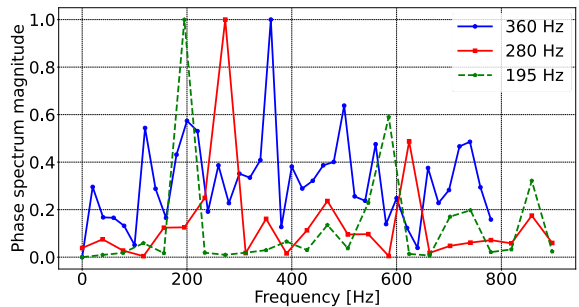

**Fig. 4:** Normalized spectra of phase difference between scattering maxima for applied acoustic signals at 195 Hz (green), 280 Hz (red), and 360 Hz (blue)

### Results obtained with 70 km fibre

In a third experiment, a 20 km standard single-mode fibre spool is added to the 50 km single-mode fibre. The reflection peak at approximately 50 km in Fig. 5 shows the connection between the two fibres. Additionally, the figure shows the reflections at the fibre end with a magnified scale in the inset, where the three reflection peaks from the connectors between the sensor fibre and the patch cord are marked. The thermal noise level is again at -103 dB, while the side lobe noise is not noticeable due to the higher loss of the signal after additionally propagating twice through 20 km of fibre. Reflections 1 and 2, separated by approximately 10 m, are used to evaluate the phase difference. Here, the probe signal frame has a length of 753 µs, and thus the maximum resolvable frequency is 664 Hz. In Fig. 6, the normalized phase spectra, obtained for four sinusoidal signals with frequencies in the range of 200 Hz to 625 Hz are shown. The frequencies of all applied signals are clearly visible, with a better signal-to-noise ratio than in Fig. 4 due to the higher reflected power.

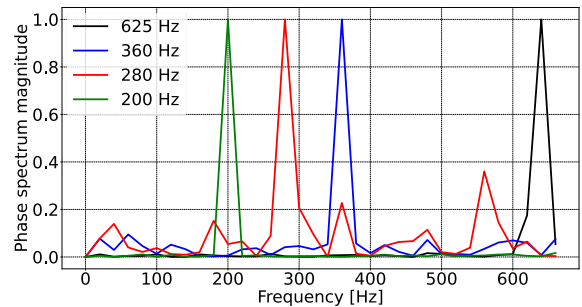

**Fig. 6:** Normalized spectra of the phase difference between reflections for applied acoustic signals at 200 Hz (green), 280 Hz (red), 360 Hz (blue), and 625 Hz (black)

### Conclusion

In this study, a COSA, designed for telecommunication applications, is integrated into a testbed with coherent correlation optical time-domain reflectometer to detect dynamic events after 50 km and 70 km of standard single-mode fibre affected by Rayleigh backscattering and Fresnel reflection. Optical phase variations due to acoustic signals with frequencies in the range from 195 Hz to 625 Hz, acting over 5 to 10 meters of fibre, were detected. This demonstrates that the telecom COSA is a suitable device for sensing applications in deployed fibre networks.

### Acknowledgements

This work has received funding from the Horizon Europe Framework Programme under grant agreement No 101093015 (SoFiN Project) and was partially funded by the German Federal Ministry of Education and Research in the framework of the RUBIN project Quantifisens (Project ID 03RU1U071D). We would like to thank NTT Electronics for providing the COSA.